\documentclass[showpacs,preprintnumbers,superscriptaddress,twocolumn]{revtex4-1}
\usepackage{amssymb}
\usepackage{amsfonts}
\usepackage{latexsym}
\usepackage{dcolumn}
\usepackage{amsmath}
\usepackage{graphicx}
\usepackage{psfrag,pstricks}
\usepackage[normalem]{ulem}
\usepackage{float}

\begin{document}
\title{Microwave Quantum Illumination}
\author{Shabir Barzanjeh}
\affiliation{Institute for
Quantum Information, RWTH Aachen University, 52056 Aachen,
Germany}
\author{Saikat Guha}
\affiliation{Quantum Information Processing Group, Raytheon BBN
Technologies, Cambridge, Massachusetts 02138, USA}
\author{Christian Weedbrook}
\affiliation{QKD Corp., 60 St. George St., Toronto, M5S 3G4,
Canada}
\author{David Vitali}
\affiliation{School of Science and Technology, University of
Camerino, Camerino, Macerata 62032, Italy}
\author{Jeffrey H. Shapiro}
\affiliation{Research Laboratory of Electronics, Massachusetts
Institute of Technology, Cambridge, Massachusetts 02139, USA}
\author{Stefano Pirandola}
\email{stefano.pirandola@york.ac.uk} \affiliation{Department of
Computer Science \& York Centre for Quantum Technologies,
University of York, York YO10 5GH, United Kingdom}

\begin{abstract}
Quantum illumination is a quantum-optical sensing technique in which an
entangled source is exploited to improve the detection of a low-reflectivity
object that is immersed in a bright thermal background. Here we describe and
analyze a system for applying this technique at microwave frequencies, a
more appropriate spectral region for target detection than the optical, due
to the naturally-occurring bright thermal background in the microwave
regime. We use an electro-opto-mechanical converter to entangle microwave
signal and optical idler fields, with the former being sent to probe the
target region and the latter being retained at the source. The microwave
radiation collected from the target region is then phase conjugated and
upconverted into an optical field that is combined with the retained idler
in a joint-detection quantum measurement. The error probability of this
microwave quantum-illumination system, or quantum radar, is shown to be
superior to that of any classical microwave radar of equal transmitted
energy.
\end{abstract}

\pacs{03.67.-a, 03.65.-w, 42.50.-p, 07.07.Df}
\maketitle

%\date{\today}

\textit{Introduction}.--~Entanglement is the foundation of many quantum
information protocols~\cite{Nielsen,Wilde,Cerf}, but it is easily destroyed
by environmental noise that, in almost all cases, kills any benefit such
nonclassical correlations would otherwise have provided. Quantum
illumination (QI)~\cite{Lloyd,pirandola}, however, is a notable exception:
it thrives in environments so noisy that they are entanglement breaking.

The original goal of QI was to detect the presence of a low-reflectivity
object, immersed in a bright thermal background, by interrogating the target
region with one optical beam while retaining its entangled counterpart for
subsequent joint measurement with the light returned from that target
region. Although the thermal noise destroys the entanglement, theory showed
that the QI system will significantly outperform a classical
(coherent-state) system of the same transmitted energy~\cite%
{pirandola,saikat,Weedbrook}. Later, a QI protocol was proposed
for secure communication~\cite{Shapiro2} whose experimental
realization~\cite{Zhang} showed that entanglement's benefit could
indeed survive an entanglement-breaking channel. Because of this
feature, QI is perhaps one of the most surprising protocols for
quantum sensing. Together with quantum
reading~\cite{Read1,Read2,Read3,Read4}, it represents a practical
example of quantum channel discrimination~\cite{Cerf}, in which
entanglement is beneficial for a technologically-driven
information task.

So far, QI has only been demonstrated at optical wavelengths~\cite%
{Zhang,Lopaeva,ZhangDetection}, for which naturally-occurring
background radiation contains far less than one photon per mode on
average, even though QI's performance advantage \emph{requires}
the presence of a bright background. The QI communication protocol
from~\cite{Shapiro2,Zhang} deals with this problem in a natural
way by purposefully injecting amplified spontaneous emission noise
to thwart eavesdropping. By contrast, similar noise injection in
QI target-detection experiments has to be considered artificial,
because better target-detection performance would be obtained
without it. The appropriate wavelengths for QI-enabled target
detection thus lie in the microwave region, in which almost all
radar systems operate and in which there is naturally-occurring
background radiation containing many photons per mode on average.
In general, the development of quantum
information techniques for microwave frequencies is quite challenging~\cite%
{Ralph,Ralph2,Ottaviani}.

In this Letter, we introduce a novel QI target-detection system that
operates in the microwave regime. Its transmitter uses an
electro-opto-mechanical (EOM) converter~\cite%
{shabir2,shabir1,lenh1,bochmann,bagci} in which a mechanical resonator
entangles signal and idler fields emitted from microwave and optical
cavities~\cite{shabir2, shabir1}. Its receiver employs another EOM
device---operating as a phase-conjugator \emph{and} a wavelength
converter---whose optical output is used in a joint measurement with the
retained idler. We show that our system dramatically outperforms a
conventional (coherent-state) microwave radar of the same transmitted
energy, achieving an orders-of-magnitude lower detection-error probability.
Moreover, our system can be realized with state-of-the-art technology, and
is suited to such potential applications as standoff sensing of
low-reflectivity objects, and environmental scanning of electrical circuits.
Thanks to its enhanced sensitivity, our system could also lead to low-flux
non-invasive techniques for protein spectroscopy and biomedical imaging.

\begin{figure}[h!]
\vspace{-0.2cm} \centering
\includegraphics[width=3.5in]{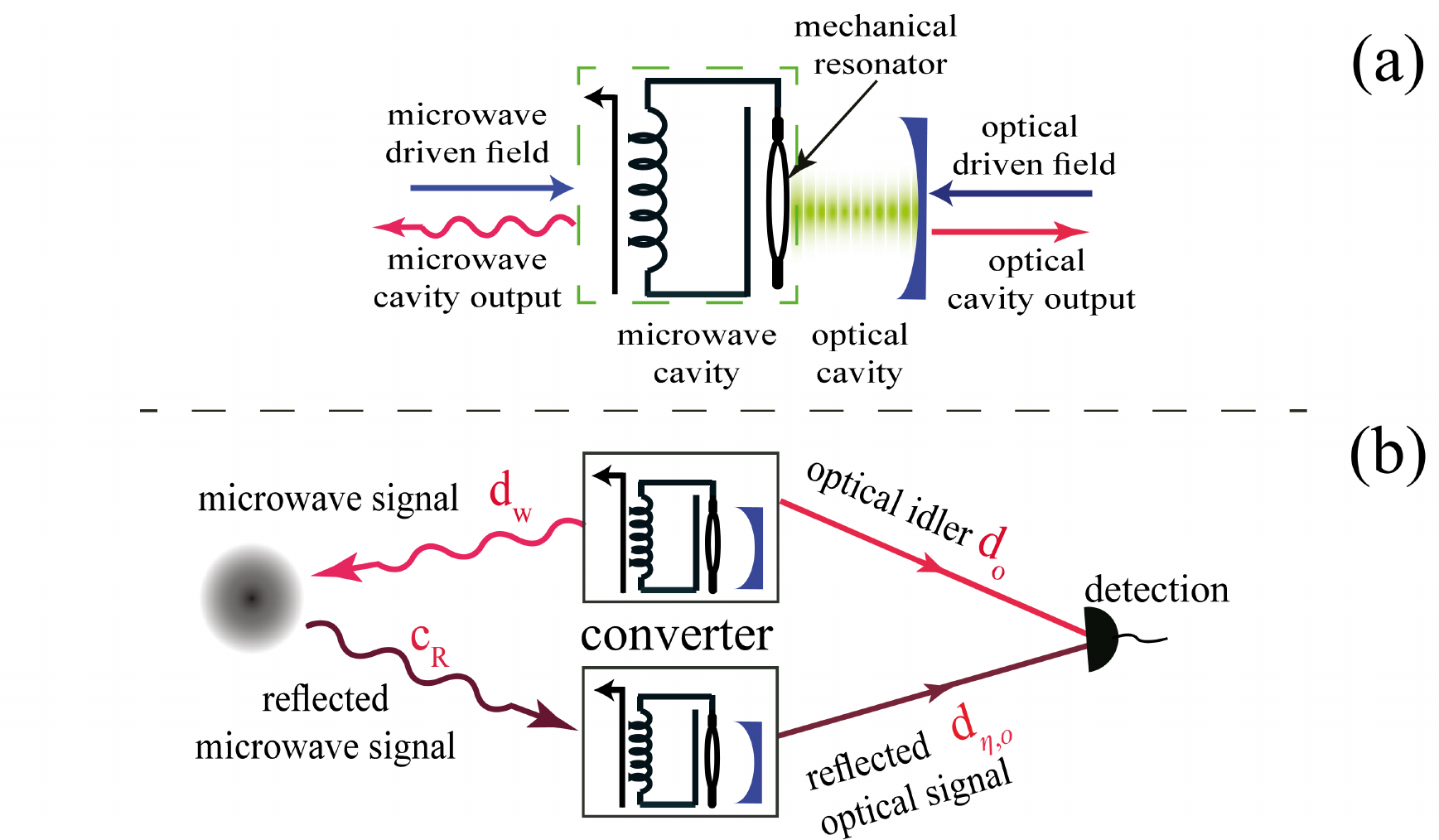}
\vspace{-0.4cm} \caption{(a) Schematic of the
electro-opto-mechanical (EOM) converter in which driven microwave
and optical cavities are coupled by a mechanical resonator. (b)
Microwave-optical QI using EOM converters. The transmitter's EOM
converter entangles microwave and optical fields. The receiver's
EOM converter transforms the returning microwave field to the
optical domain while performing a phase-conjugate operation.}
\label{fig1}
\end{figure}

\textit{Electro-opto-mechanical converter}.--~As depicted in Fig.~\ref{fig1}%
(a), the EOM converter couples a microwave-cavity mode (annihilation
operator $\hat{a}_{\text{w}}$, frequency $\omega _{\text{w}}$, damping rate $%
\kappa _{\text{w}}$) to an optical-cavity mode (operator $\hat{a}_{o}$,
frequency $\omega _{o}$, damping rate $\kappa _{o}$) through a mechanical
resonator (operator $\hat{b}$, frequency $\omega _{M}$, damping rate $\gamma
_{M}$) \cite{shabir1,lenh1, bochmann}. In the frame rotating at the
frequencies of the microwave and optical driving fields, the interaction
between the cavities' photons and the resonator's phonons is governed by the
Hamiltonian \cite{supp}
\begin{equation*}
\hat{H}=\hbar \omega _{M}\hat{b}^{\dagger }\hat{b}+\hbar \sum_{j=\text{w},o}%
\left[ \Delta _{0,j}+g_{j}(\hat{b}+\hat{b}^{\dagger })\right] \hat{a}%
_{j}^{\dagger }\hat{a}_{j}+\hat{H}_{\text{dri}}.
\end{equation*}%
Here, $g_{j}$ is the coupling rate between the mechanical resonator and
cavity $j$, which is driven at frequency $\omega _{j}-\Delta _{0,j}$ by the
coherent-driving Hamiltonian $\hat{H}_{\text{dri}}$~\cite{supp}.

The electro-opto-mechanical coupling rates $g_j$ are typically small, so
that we can linearize the Hamiltonian by expanding the cavity modes around
their steady-state field amplitudes $\hat{c}_{j}=\hat{a}_{j}-\sqrt{N_{j}}$,
where the $N_{j}\gg1$ are the mean numbers of cavity photons induced by the
pumps~\cite{shabir2, lenh}. In the interaction picture with respect to the
free Hamiltonian, we may then write~\cite{supp}%
\begin{equation}
\hat{H}=\hbar G_{o}(\hat{c}_{o}\hat{b}+\hat{b}^{\dagger}\hat{c}_{o}^{\dagger
})+\hbar G_{\mathrm{w}}(\hat{c}_{\mathrm{w}}\hat{b}^{\dagger}+\hat{b}\hat {c}%
_{\mathrm{w}}^{\dagger}),  \label{hameff}
\end{equation}
where $G_{j}=g_{j}\sqrt{N_j}$ is the multi-photon coupling rate. This
expression assumes that the effective cavity detunings satisfy $\Delta_{%
\mathrm{w}}=-\Delta_{o}=\omega_{M}$ and that resonator is in its
fast-oscillation regime, so that the red sideband drives the microwave
cavity while the blue sideband drives the optical cavity and we can neglect
terms oscillating at $\pm 2\omega_{M}$.

Equation~(\ref{hameff}) shows that the mechanical resonator
mediates a delayed interaction between the optical and microwave
cavity modes. Its first term is a parametric down-conversion
interaction that entangles the mechanical resonator and the
optical cavity mode. This entanglement is transmitted to the
propagating optical mode $\hat{d}_{o}$, if the opto-mechanical
rate $G_{o}^{2}/\kappa _{o}$ exceeds the decoherence rate of the
mechanical resonator $r=\gamma_{M}\bar{n}_{b}^{T}\approx
\gamma_{M}k_{B}T_{\text{EOM}}/\hbar \omega _{M}$, where $k_{B}$ is
Boltzmann's constant, $T_{\text{EOM}}$ is the EOM converter's
absolute temperature, $\bar{n}_{b}^{T}=[\mathrm{e}^{\hbar \omega
_{M}/(k_{B}T_{\text{EOM}})}-1]^{-1}$, and the approximation
presumes $k_{B}T_{\text{EOM}}\ \gg \hbar \omega _{M}$, as will
be the case for the parameter values assumed later. The second term in Eq.~(%
\ref{hameff}) is a beam-splitter interaction between the mechanical
resonator and the microwave cavity mode that transfers the entanglement to
the propagating microwave field $\hat{d}_{\text{w}}$, as long as the
microwave-mechanical rate satisfies $G_{\mathrm{w}}^{2}/\kappa _{\mathrm{w}%
}>r$~\cite{lenh,hofer}.

\textit{Microwave-optical entanglement}.--~The output propagating modes can
be expressed in terms of the intracavity quantum noise operators, $\hat{c}%
_{j,\mathrm{in}}$, and the quantum Brownian noise operator, $\hat{b}_{%
\mathrm{int}}$, via~\cite{supp}
\begin{align}
\hat{d}_{\mathrm{w}}& =A_{\mathrm{w}}\hat{c}_{\mathrm{w},\mathrm{in}}-B\hat{c%
}_{o,\mathrm{in}}^{\dagger }-C_{\mathrm{w}}\hat{b}_{\text{int}},
\label{qle1} \\
\hat{d}_{o}& =B\hat{c}_{{\mathrm{w}},\mathrm{in}}^{\dagger }+A_{o}\hat{c}_{o,%
\mathrm{in}}-C_{o}\hat{b}_{\text{int}}^{\dagger },  \label{downconverter2}
\end{align}%
where $A_{j}$, $B$, and $C_{j}$ depend on the cooperativity terms $\Gamma
_{j}=G_{j}^{2}/\kappa _{j}\gamma _{M}$ as given in~\cite{supp}. The $\hat{c}%
_{\mathrm{w},\mathrm{in}},\hat{c}_{\mathrm{o,in}}$ and $\hat{b}_{\text{int}}$
modes in Eqs.~(\ref{qle1}) and (\ref{downconverter2}) are in independent
thermal states whose average photon numbers, $\bar{n}_{\text{w}}^{T}$, $\bar{%
n}_{o}^{T}$, and $\bar{n}_{b}^{T}$, are given by temperature-$T_{\text{EOM}}$
Planck laws at their respective frequencies. It follows that the propagating
modes, $\hat{d}_{\text{w}}$ and $\hat{d}_{o}$, are in a zero-mean,
jointly-Gaussian state completely characterized by the second moments,
\begin{align*}
\bar{n}_{\text{w}}& \equiv \langle \hat{d}_{\text{w}}^{\dagger }\hat{d}_{%
\text{w}}\rangle =|A_{\text{w}}|^{2}\bar{n}_{\text{w}}^{T}+|B|^{2}(\bar{n}%
_{o}^{T}+1)+|C_{\text{w}}|^{2}\bar{n}_{b}^{T}, \\
\bar{n}_{o}& \equiv \langle \hat{d}_{o}^{\dagger }\hat{d}_{o}\rangle
=|B|^{2}(\bar{n}_{\text{w}}^{T}+1)+|A_{o}|^{2}\bar{n}_{o}^{T}+|C_{o}|^{2}(%
\bar{n}_{b}^{T}+1), \\
\langle \hat{d}_{\text{w}}\hat{d}_{o}\rangle & =A_{\text{w}}B(\bar{n}_{\text{%
w}}^{T}+1)-BA_{o}\bar{n}_{o}^{T}+C_{\text{w}}C_{o}(\bar{n}_{b}^{T}+1).
\end{align*}

\begin{figure}[h]
\vspace{-0.1cm} \centering
\includegraphics[width=3.52in]{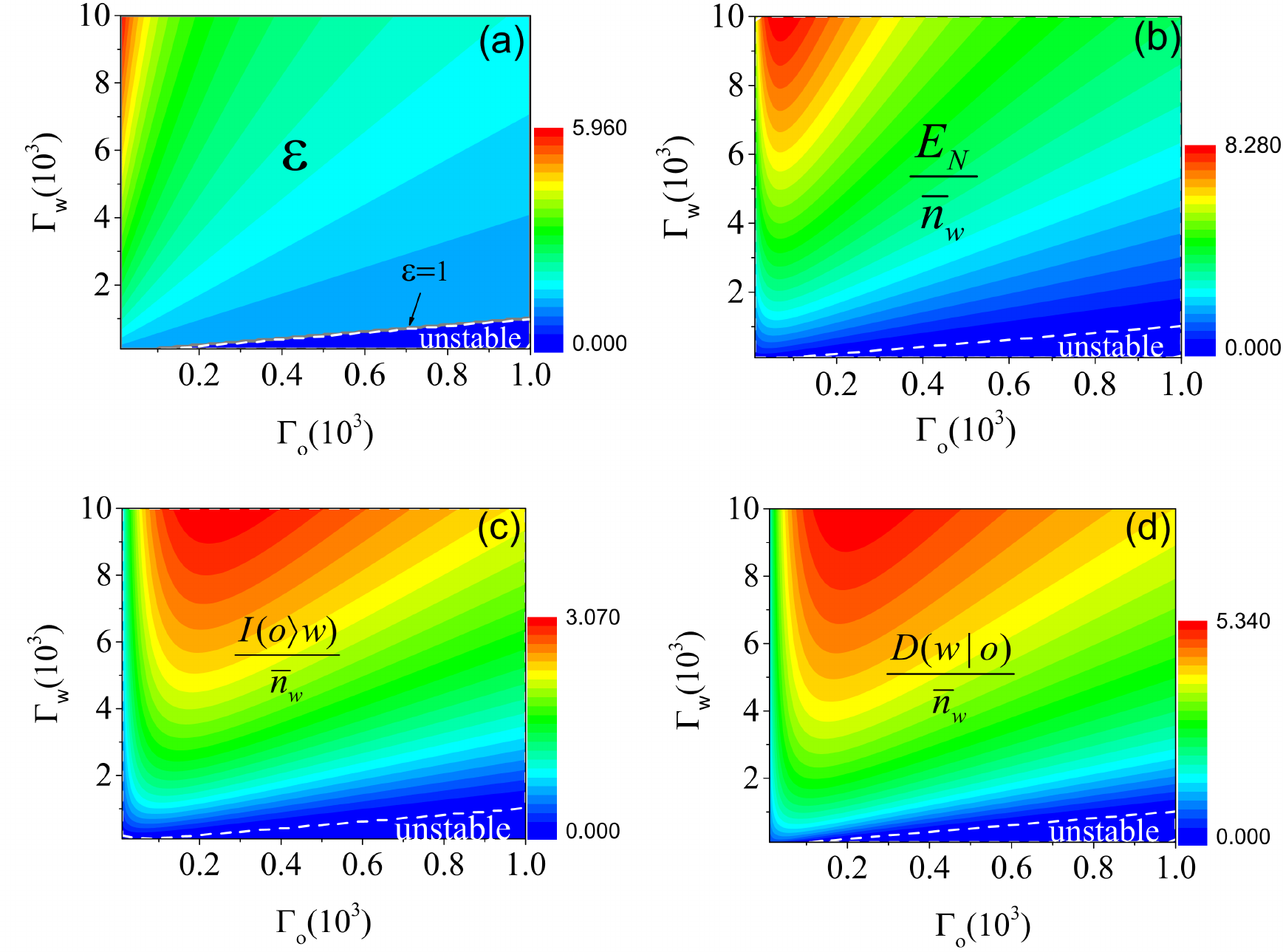}
\vspace{-0.5cm} \caption{Performance of our microwave-optical
source versus the cooperativity parameters $\Gamma _{\text{w}}$
and $\Gamma _{o}$. We show the behaviour of the entanglement
metric $\mathcal{E}$ (abstract units) in panel (a), the normalized
logarithmic negativity $E_{N}/\bar{n}_{\text{w}}$ (ebits
per microwave photon) in panel (b), the normalized coherent information $%
I(o\rangle $w$)/\bar{n}_{\text{w}}$ (qubits per microwave photon) in panel
(c), and the normalized quantum discord $D({\text{w}}|o)/\bar{n}_{\text{w}}$
(discordant bits per microwave photon) in panel (d). In all panels we assume
experimentally-achievable parameters~\protect\cite{lenh,palo}: a 10-ng-mass
mechanical resonator with $\protect\omega _{M}/2\protect\pi =10$\thinspace
MHz and $Q=30\times 10^{3}$; a microwave cavity with $\protect\omega _{%
\mathrm{w}}/2\protect\pi =10$\thinspace GHz and $\protect\kappa _{\mathrm{w}%
}=0.2\,\protect\omega _{M}$; and a 1-mm-long optical cavity with $\protect%
\kappa _{o}=0.1\,\protect\omega _{M}$ driven by a 1064-nm-wavelength laser.
The opto-mechanical and electro-mechanical coupling rates are $g_{\mathrm{o}%
}/2\protect\pi =115.512$\thinspace Hz and $g_{\mathrm{w}}/2\protect\pi %
=0.327 $\thinspace Hz, and the entire EOM converter is held at temperature $%
T_{\text{EOM}}\ =30$\thinspace mK. In each panel, the boundary between
stable and unstable operation was obtained from the Routh-Hurwitz criterion~%
\protect\cite{Stability}.}
\label{fig2}
\end{figure}

The propagating microwave and optical fields will be entangled if and only
if the metric $\mathcal{E}\equiv |\langle \hat{d}_{\text{w}}\hat{d}%
_{o}\rangle |/\sqrt{\bar{n}_{\text{w}}\bar{n}_{o}}$ is greater than $1$~\cite%
{supp}. As we can see from Fig.~\ref{fig2}, there is a wide region where $%
\mathcal{E}>1$ in the plane of the cooperativity parameters, $\Gamma _{\text{%
w}}$ and $\Gamma _{o}$, varied by varying the microwave and
optical powers driving their respective cavities, and assuming
experimentally-achievable system parameters~\cite{lenh, palo}. The
threshold condition $\mathcal{E}=1$ almost coincides with the
boundary between the stable and unstable parameter regions, as
given by the Routh-Hurwitz criterion~\cite{Stability}.

The quality of our microwave-optical source can also be evaluated using
measures of quantum correlations, as typical in quantum information. Since
the QI's advantage is computed at fixed mean number of microwave photons $%
\bar{n}_{\text{w}}$ irradiated through the target, the most powerful quantum
resources are expected to be those maximizing their quantum correlations per
microwave photon emitted. Following this physical intuition, we analyze our
source in terms of the normalized log-negativity$~$\cite{eis} $E_{N}/\bar{n}%
_{\text{w}}$ and the normalized coherent information~\cite{coh1,coh2} $%
I(o\rangle $w$)/\bar{n}_{\text{w}}$. Respectively, they represent an upper
and a lower bound to the mean number of entanglement bits (ebits) which are
distillable for each microwave photon emitted by the source~\cite{supp}.
Furthermore, since our source is in a mixed state (more precisely, a
two-mode squeezed thermal state), we also quantify its normalized quantum
discord~\cite{Discord,supp} $D({\text{w}}|o)/\bar{n}_{\text{w}}$, which
captures the quantum correlations carried by each microwave photon. As we
can see from Fig.~\ref{fig2}, our source has a remarkable performance in
producing distillable ebits and discordant bits for each microwave photon
emitted.

\textit{Microwave quantum illumination}.--~For QI target
detection, our signal-idler mode pair analysis must be extended to
a continuous-wave EOM converter whose
$W_m$-Hz-bandwidth~\cite{bandwidth} output fields are used in a
$t_m$-sec-duration measurement involving $M = t_mW_m \gg 1$
independent, identically-distributed (iid) mode pairs. The $M$
signal modes
interrogate the target region that is equally likely to contain (hypothesis $%
H_1$) or not contain (hypothesis $H_0$) a low-reflectivity object. Either
way, the microwave field that is returned consists of $M$ iid modes. Using $%
\hat{c}_R$ to denote the annihilation operator for the mode returned from
transmission of $\hat{d}_\text{w}$, we have that $\hat{c}_R = \hat{c}_B$
under hypothesis $H_0$, and $\hat{c}_R = \sqrt{\eta}\,\hat{d}_\text{w} +
\sqrt{1-\eta}\,\hat{c}_B$, under hypothesis $H_1$. Here, $0 < \eta \ll 1$ is
the roundtrip transmitter-to-target-to-receiver transmissivity (including
propagation losses and target reflectivity), and the background-noise mode, $%
\hat{c}_B$, is in a thermal state with temperature-$T_B$ Planck-law average
photon number $\bar{n}_{B}$ under $H_0$, and in a thermal state with $\bar{n}%
_{B}/(1-\eta) \approx \bar{n}_B$ under $H_1$~\cite{pirandola}. See
Fig.~\ref{fig1}(b).

Under $H_{1}$, the returned microwave and the retained optical fields are in
a zero-mean, jointly-Gaussian state with a nonzero phase-sensitive cross
correlation $\langle \hat{c}_{R}\hat{d}_{o}\rangle $ that is invariant to
the $\bar{n}_{B}$ value, while $\langle \hat{c}_{R}^{\dagger }\hat{c}%
_{R}\rangle $ increases with increasing $\bar{n}_{B}$. Consequently, the
returned and retained radiation under $H_{1}$ will \emph{not} be entangled
when
\begin{equation*}
\bar{n}_{B}\geq \bar{n}_{B}^{\text{thresh}}\equiv \eta \left( |\langle \hat{d%
}_{\text{w}}\hat{d}_{o}\rangle |^{2}/\bar{n}_{o}-\bar{n}_{\text{w}}\right) .
\end{equation*}

\textit{Microwave-to-optical phase-conjugate receiver}.--~The receiver
passes the $M$ return modes into the microwave cavity of another (identical)
EOM converter to produce $M$ iid optical-output modes each given by $\hat{d}%
_{\eta ,\mathrm{o}}=B\hat{c}_{R}^{\dagger }+A_{\mathrm{o}}\hat{c}_{\mathrm{%
o,in}}^{\prime }-C_{\mathrm{o}}\hat{b}_{\text{int}}^{^{\prime }\dagger }$,
where $\{\hat{c}_{\text{w,in}}^{\prime },\hat{c}_{o,\text{in}}^{\prime },%
\hat{b}_{\text{int}}^{\prime }\}$ have the same states as their counterparts
in the transmitter's EOM converter. The receiver's EOM converter thus phase
conjugates the returned microwave field \emph{and} upconverts it to an
optical field. This output is combined with the retained idler on a 50--50
beam splitter whose outputs are photodetected and their photon counts---over
the $t_{m}$-sec-long measurement interval---are subtracted to yield an
outcome from which a minimum error-probability decision about object absence
or presence will be made~\cite{saikat}. For $M\gg 1$, the resulting error
probability is~\cite{supp,saikat} $P_{\text{QI}}^{(M)}=\mathrm{erfc}\left(
\sqrt{\text{SNR}_{\text{QI}}^{(M)}/8}\,\right) /2,$ with SNR$_{\text{QI}%
}^{(M)}$ being the QI system's signal-to-noise ratio for its $M$
mode pairs~\cite{supp} and erfc(...) being the complementary error
function~\cite{saikat}.

\textit{Comparison with classical microwave transmitters}.-- Suppose that a
coherent-state microwave transmitter---emitting $M\bar{n}_{\text{w}}$
photons on average, with $\bar{n}_{\text{w}}$ equaling the mean number of
microwave photons per mode emitted by our source---is used to assess target
absence or presence. Homodyne detection of the microwave field returned from
the target region followed by minimum error-probability processing of its
output results in an error probability~\cite{saikat} $P_{\text{coh}}^{(M)}=%
\mathrm{erfc}\!\left( \sqrt{\text{SNR}_{\text{coh}}^{(M)}/8}\,\right) \!/2$,
in terms of this system's signal-to-noise ratio, $\text{SNR}_{\text{coh}%
}^{(M)}=4\eta M\bar{n}_{\text{w}}/(2\bar{n}_{B}+1)$. This performance
approximates the error exponent of the quantum Chernoff bound~\cite%
{QCB1,QCB2,QCBgauss} computed for $M\gg 1$, implying that homodyne detection
is the asymptotically optimal receiver for target detection when a
coherent-state transmitter is employed.

\begin{figure}[h!]
\vspace{-0.1cm} \centering
\includegraphics[width=2.5in]{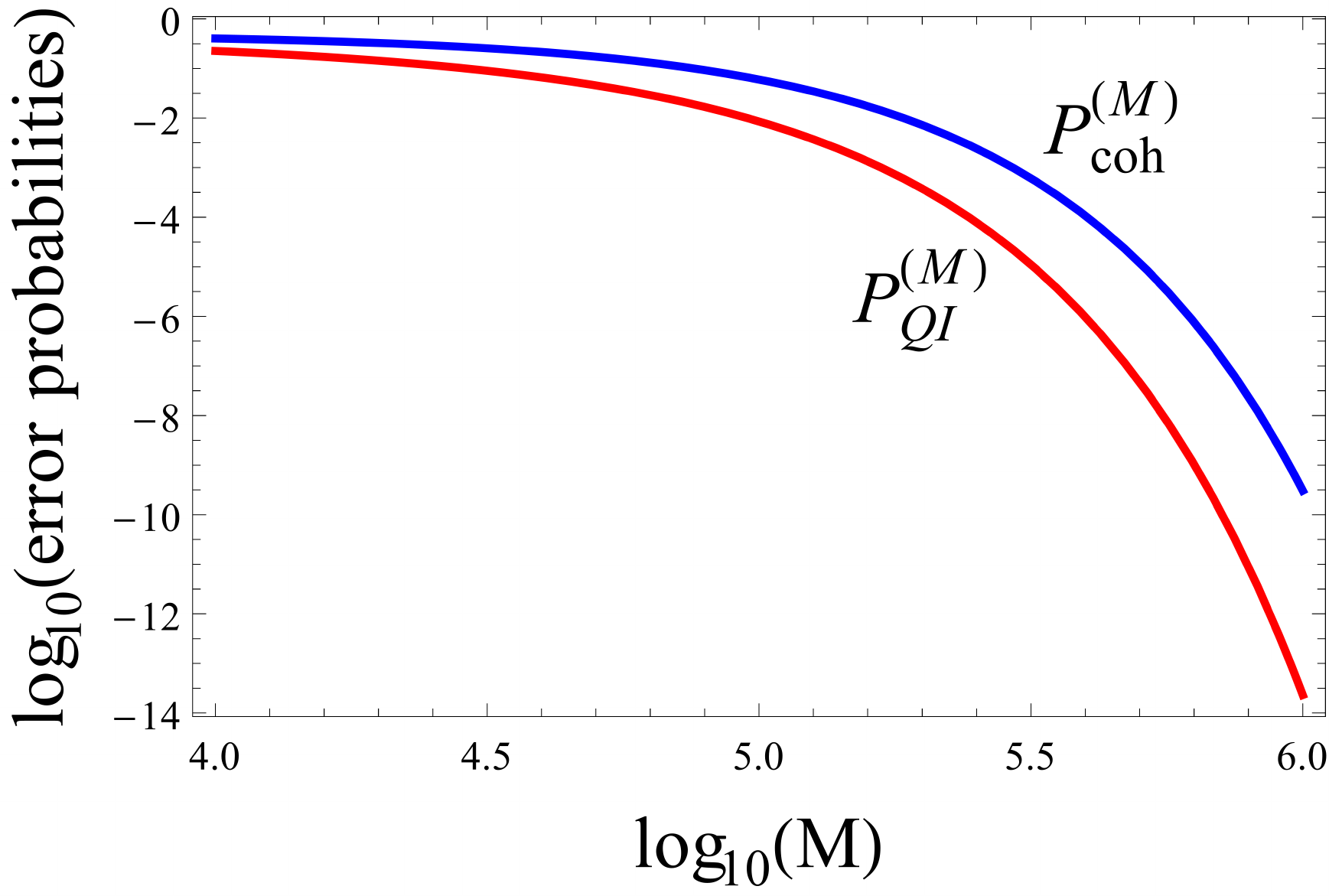}
\vspace{-0.1cm} \caption{$P_{\text{QI}}^{(M)}$ and
$P_{\text{coh}}^{(M)}$\ versus the
time-bandwidth product, $M$, assuming $\protect\eta =0.07$, $\Gamma _{\text{w%
}}\ =5181.95$ and $\Gamma _{o}=668.43$ (implying $\bar{n}_{\text{w}}\ =0.739$
and $\bar{n}_{o}=0.681$), and room temperature $T_{B}=293\,$K (implying $%
\bar{n}_{B}=610\gg \bar{n}_{B}^{\text{thresh}}=0.069$).}
\label{QCB}
\end{figure}

Figure~\ref{QCB} plots $P_{\text{QI}}^{(M)}$ and $P_{\text{coh}}^{(M)}$
versus $\log _{10}M$ for the EOM converter parameters given in Fig.~\ref%
{fig2} and $\eta =0.07$. It assumes $\Gamma _{\text{w}}\ =5181.95$ and $%
\Gamma _{o}=668.43$ (implying $\bar{n}_{\text{w}}=0.739$ and $\bar{n}%
_{o}=0.681$) and $T_{B}=293\,$K (implying $\bar{n}_{B}=610$). We
see that the QI system can have an error probability that is
orders of magnitude lower than that of the coherent-state system.
Moreover, according to Ref.~\cite{pirandola}, no other
classical-state system with the same energy constraint can have a
lower error probability than the coherent-state system.

To further study the performance gain of our microwave QI system over a
classical sensor, we evaluate $\mathcal{F}\equiv \text{SNR}_{\text{QI}%
}^{(M)}/\text{SNR}_{\text{coh}}^{(M)}$ for large $M$. This figure of merit
depends on the cooperativity parameters, $\Gamma _{\text{w}}$ and $\Gamma
_{o}$, whose values are typically large $\Gamma _{j}\gg 1$ (cf.\ the values
in Fig.~\ref{fig2}, which rely on experimentally-achievable parameters) and
the brightness of the background, $\bar{n}_{B}$. As shown in Fig.~\ref%
{qicrit}, QI's superiority prevails in a substantial region of $\Gamma _{%
\text{w}}$, $\Gamma _{o}$ values corresponding to Fig.~\ref{fig2} regions
where our source has the best efficiency in producing quantum entanglement
and, more generally, quantum correlations, per microwave photon emitted.
Such advantage is found as long as the average number of microwave photons
is sufficiently low.
\begin{figure}[h!]
\vspace{-0.7cm} \centering
\includegraphics[width=2.9in]{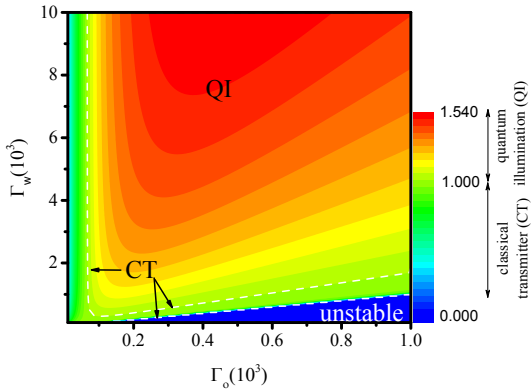}
\vspace{-0.3cm}
\caption{QI-advantage figure of merit, $\mathcal{F}$, versus $\Gamma _{%
\mathrm{w}}$ and $\Gamma _{o}$. For $\mathcal{F}>1$, the QI system
has lower error probability than any classical-state system, i.e.,
classical transmitter (CT), of the same transmitted energy. See
Fig. \protect\ref{fig2} for the other parameter values.}
\label{qicrit}
\end{figure}

\textit{Conclusion and Discussion}.--~We have shown that quantum
illumination can be performed in its more natural setting for
target detection, i.e., the microwave regime, by using a pair of
electro-opto-mechanical converters. Thanks to this converter, the
target region can be interrogated at a microwave frequency, while
the quantum-illumination joint measurement needed for target
detection is made at optical frequency, where the high-performance
photodetectors needed to obtain QI's performance advantage are
available.

An optimized EOM converter is able to generate strong quantum entanglement
and quantum correlations between its microwave and optical outputs. These
correlations can successfully be exploited for target detection, yielding
lower error probability than that of any classical-state microwave system of
the same transmitted energy. The QI advantage is especially evident when
detecting the faint returns from low-reflectivity objects embedded in the
bright thermal noise typical of room-temperature microwave environments.

%%%%%%%%%%%%%%%%%%%%%%%%%%

Note that we assumed unit quantum efficiency for the optical part
of our quantum receiver. This is not far from current experimental
conditions: photon collection efficiencies from optical cavities
can be very high ($>74\%$ in Ref.~\cite{Rempe13}), loss at the
beam splitter can be extremely low, and photodetection can be
extremely efficient at optical wavelengths. Thus the main source
of loss may come from the optical storage of the idler mode, to be
preserved during the signal roundtrip time. This is not an issue
for short-range applications but, for long-range tasks, the idler
loss must remain below 3~dB, otherwise the QI advantage of the
phase-conjugating quantum receiver is lost~\cite{saikat}. While
using a good quantum memory (e.g., a rare-earth
doped-crystal~\cite{Zhong15}) would solve the problem, the
practical solution of storing the idler into an optical-fiber
delay line would restrict the maximum range of the quantum radar
to about $11.25$~km in free-space (assuming a fiber loss of
$0.2$~dB/km and fiber propagation speed equal to $2c/3$, where $c$
is vacuum light-speed).

%%%%%%%%%%%%%%%%%%%%%%%%

Finally, extending our results to lower frequencies (below
1\thinspace GHz), our scheme could potentially be used for
non-invasive NMR spectroscopy in structural biology (structure of
proteins and nucleic acids) and in medical applications (magnetic
resonance imaging). Future implementations of quantum illumination
at the microwave regime could also be achieved by using other
quantum sources, for instance based on Josephson parametric
amplifiers, which are able to generate entangled microwave modes
of high quality~\cite{JPA1,JPA2,JPA3,JPA4}. These amplifiers might
become a very good choice once that suitable high-performance
microwave photo-detectors are made available.

\textit{Acknowledgments}.--~S.B. is grateful for support from the
Alexander von Humboldt foundation. D.V. is sponsored by the
European Commission (ITN-Marie Curie Project \lq cQOM\rq, Grant
No. 290161, and FET-Open Project \lq iQUOEMS\rq, Grant No.
323924). S.G. was supported by the US Office of Naval Research
contract number N00014-14-C-0002. J.H.S. appreciates sponsorship
by AFOSR and ONR. S.P. has been sponsored by a Leverhulme Trust
research fellowship (\lq qBIO\rq ) and EPSRC via \lq
qDATA\rq~(Grant no. EP/L011298/1) and \lq HIPERCOM\rq~(Grant No.
EP/J00796X/1).


\begin{thebibliography}{99}
\bibitem{bandwidth} The common bandwidth of the entangled microwave and
optical fields is $W_m = \gamma_M(\Gamma_\text{w}-\Gamma_o + 1)$, in terms
of the resonator's damping rate and the cooperativity parameters.
\bibitem{Stability} The system is stable if the Routh-Hurwitz criterion is
satisfied. For $\Gamma _{i}\gg 0$, this criterion reduces to the following
necessary and sufficient condition~\cite{Wang}:
\begin{equation*}
\kappa _{\text{w}}\Gamma _{\text{w}}-\kappa _{o}\Gamma _{o}>\tilde{K}\,\text{max}\left\{ \kappa _{\text{w}}-\kappa _{o},\frac{\kappa _{o}^{2}-\kappa _{\text{w}}^{2}}{2\gamma _{M}+\kappa _{\text{w}%
}+\kappa _{o}}\right\} ,
\end{equation*}%
where $\tilde{K}=\Gamma _{\text{w}}(1+\kappa
_{o}/\kappa_{\text{w}})^{-1}+\Gamma _{o}(1+\kappa
_{\text{w}}/\kappa _{o})^{-1}$.
\end{thebibliography}
\end{document}